\documentclass[12pt]{article}
\usepackage{amssymb}
\usepackage{amsfonts}
\usepackage{amsmath}

\def\sb{{\sf b}}

\def\cH{{\cal H}}

\newcommand\ket[1]{| #1 \rangle}

\newcommand\ketbra[2]{|#1\rangle\langle #2|}

\newcommand\trnorm[1]{\| #1 \|_1}
\catcode`@=11  
\def\@cite#1{$^{#1}$}
\def\@refe#1{[#1]}
\def\@biblabel#1{{\normalsize\bf{#1}}}
\def\refe{\@ifnextchar [{\@tempswatrue\@citexr}{\@tempswafalse\@citexr[]}}
\def\@citexr[#1]#2{\if@filesw\immediate\write\@auxout{\string\citation{#2}}\fi
  \def\@citea{}\@refe{\@for\@citeb:=#2\do
    {\@citea\def\@citea{,}\@ifundefined
       {b@\@citeb}{{\bf ?}\@warning
       {Citation `\@citeb' on page \thepage \space undefined}}%
\hbox{\csname b@\@citeb\endcsname}}}{#1}}
 \catcode`@=12

\begin{document}

\title{\bf Why there is no impossibility theorem on  Secure Quantum Bit Commitment}

\author{Horace P. Yuen\\
Department of Electrical and Computer Engineering\\
Department
of Physics and Astronomy\\
Northwestern University, Evanston, IL
60208-3118, USA\\
E-mail: yuen@ece.northwestern.edu}

\date{}

\maketitle

\begin{abstract}
The impossibility proof on unconditionally secure quantum
bit commitment is critically reviewed. Different ways of obtaining
secure protocols are indicated.
\end{abstract}

\vfill

\noindent{NOTE: This article is going to appear in the 2002 QCMC
Proceedings, and is based on quant-ph/0207089.  It contains a concise
summary of several gaps in the QBC impossibility proof, and a brief
description of an unconditionally secure protocol {\bf QBC1}.  Of all
the QBC protocols I have been presenting so far with various claims, I
will in the not-too-distant future elaborate on which ones are secure
as they are, which ones can be modified to be secure, which ones (such
as {\bf QBC4}) are essentially insecure, and which ones have undecided
security status.  This should clarify and correct any ambiguous or
erroneous statements concerning these protocols.}

\newpage

\baselineskip 20pt

\section{Introduction}\label{sec:intro}

There is a nearly universal acceptance of the general
impossibility\cite{may1}$^{\rm -}$\cite{gl} of
secure quantum bit commitment (QBC), taken to be a consequence of the
Einstein-Podolsky-Rosen (EPR) type entanglement cheating which rules out QBC and other quantum protocols that have been
proposed for various cryptographic objectives. Since there is no
characterization of all possible QBC protocols, logically there can be
no general impossibility proof as maintained to this date.
In this article, which is based on Ref.~\refe{yue5}, we explain the
nature of various gaps and incompleteness in the impossibility proof,
in addition to this a priori logical point. 
They should make clear the fact that there is no impossibility theorem
even in the absence of a specific protocol that has been proved
unconditionally secure. But we also describe an unconditionally secure
protocol {\bf QBC1} and other possible approaches for obtaining secure
protocols.

\section{The impossibility proof}\label{sec:impproof}

The essential ideas that 
constitute the impossibility proof are generally agreed upon.\cite{may1}$^{\rm -}$\cite{gl}  Adam and Babe have available to them two-way quantum
communications that terminate in a finite number of exchanges, during
which either party can perform any operation allowed by the laws of
quantum physics.  During these exchanges, Adam would have
committed a bit with associated evidence to Babe.  It is argued that,
at the end of the commitment phase, there is an entangled pure state
$\ket{\Phi_\sb}$, $\sb \in \{0,1\}$, shared between Adam who
possesses state space $\cH^A$, and Babe who possesses $\cH^B$.  For
example, if Adam sends Babe one of $M$ possible states $\{
\ket{\phi_{\sb i}} \}$ for bit \sb\ with probability $p_{\sb i}$, then
$\ket{\Phi_{\sb }} = \sum_i \sqrt{p_{\sb i}}\ket{e_i}\ket{\phi_{\sb
i}}$ with orthonormal $\ket{e_i} \in \cH^A$ and given $\ket{\phi_{\sb i}}
\in \cH^B$.  Adam would open by making a measurement on $\cH^A$, say
$\{ \ket{e_i} \}$, communicating to Babe his result $i_0$ and $\sb$;
then Babe would verify by measuring
 $\ketbra{\phi_{\sb i_0}}{\phi_{\sb i_0}}$ on $\cH^B$,
accepting as correct only the result 1.

Generally, Babe can try to identify the bit from $\rho^B_\sb$, the
marginal state of $\ket{\Phi_\sb}$ on $\cH^B$, by performing an
optimal quantum measurement that yields the optimal cheating
probability $\bar{P}^B_c$ for her.  Adam cheats by committing
$\ket{\Phi_0}$ and making a measurement on $\cH^A$ to open $i_0$ and
$\sb=1$.  His probability of successful cheating is computed through
$\ket{\Phi_\sb}$, his particular measurement, and Babe's verifying
measurement; the one optimized over all of his possible actions will
be denoted $\bar{P}^A_c$.  For a fixed measurement basis, Adam's
cheating can be described by a unitary operator $U^A$ on $\cH^A$. When
$\rho^B_0 = \rho^B_1$, i.e., $\bar{P}^B_c = 1/2$, $U^A$ is obtained
via the Schmidt decomposition of $\ket{\Phi_\sb}$. For unconditional, rather than perfect, security, one demands that
both cheating probabilities $\bar{P}^B_c - 1/2$ and $\bar{P}^A_c$ can
be made arbitarily small when a security parameter $n$ is increased.\cite{may2} Thus, {\it unconditional security} is quantitatively expressed
as
\begin{equation}
({\rm US}) \qquad \lim_n \bar{P}^B_c = \frac{1}{2},\quad \lim_n
  \bar{P}^A_c = 0.
\label{eq:us}
\end{equation}
This condition (\ref{eq:us}) says that, for any $\epsilon > 0$, there
exists an $n_0$ such that for all $n > n_0$, $\bar{P}^B_c - 1/2 \le
\epsilon$ and $\bar{P}^A_c \le \epsilon$, to which we  refer as
$\epsilon$-{\it concealing} and $\epsilon$-{\it binding}.  These
cheating probabilities are to be computed purely on the basis of
physical laws, and thus would survive any change in
technology, including any increase in computational power.  One can write down explicitly $\bar{P}^B_c = \frac{1}{4}\left(2 + \trnorm{\rho^B_0 -
  \rho^B_1}\right)$. The corresponding $\bar{P}^A_c$ 
satisfies:\cite{yue5,yue4}
\begin{equation}
4 (1-\bar{P}^B_c)^2 \le \bar{P}^A_c \le 2\sqrt{\bar{P}^B_c (1-\bar{P}^B_c)}.
\label{eq:fidbound}
\end{equation}
The lower bound in (\ref{eq:fidbound}) yields the impossibility proof \cite{may1,yue2} 
\begin{equation}
({\rm IP}) \qquad \lim_n \bar{P}^B_c = \frac{1}{2} \,\, \Rightarrow
  \,\, \lim_n \bar{P}^A_c = 1
\label{eq:ip}
\end{equation}
\par When random numbers known only to one party are used in
the commitment, they are to be replaced by corresponding 
entanglement purification. For a random $k$, it is argued
from the doctrine of the ``Church of the Larger Hilbert
Space''\cite{gl} that it is to be replaced by the purification
$|\Psi\rangle$ in ${\cal H}^{B_1}\otimes{\cal H}^{B_2}$,
\begin{equation}
|\Psi\rangle=\sum_k\sqrt{\lambda_k}|\psi_k\rangle|f_k\rangle,
\label{Psi}
\end{equation}
where the $|f_k\rangle$'s
are complete orthonormal in ${\cal H}^{B_2}$ kept by Babe while
${\cal H}^{B_1}$ would be sent to Adam. Similar purification is to be
used for performing any operation during commitment that might
otherwise require an actual measurement. As a consequence,
it is claimed that a shared state $|\Phi_\sb\rangle$ at the end of commitment is known to both parties. 
\par It appears that there are many incompleteness in the impossibility
proof. For
example, one may observe that the cheating probability
$\bar{P}^A_c$ depends on Babe's verifying measurement. For an
arbitrary protocol, the impossibility proof formulation does not, and
in fact, cannot specify what the possible verifying measurements could be. There is
{\em no} proof given that there cannot be more than one verifying
measurement for which different cheating transformations are
needed. When such a situation occurs, Adam may not know which one to
use for a successful cheating. Even though this gap can be closed, in a
proof that is not totally obvious, it is indicative of the
incompleteness of the impossibility proof. The followin situations
show that the impossibility proof formulation is actually widely
incomplete. A protocol may involve cheating detection during
commitment with corresponding possibility of aborting the protocol, a situation
different from cheat-sensitive protocols\cite{Hardy}. It has
to be decided what would happen when cheating is detected, say in a
game-theoretic formulation. It makes no sense to keep trying until
one party's cheating is not detected; some limit on the number of
detected cheats must be imposed. Assuming both parties are
honest not trying to cheat, which is what the impossibility proof
formulation does except for Adam to form entanglement instead of
sending one $|\phi_{\text{b}i}\rangle$, also makes no sense because
there would then be no need for a protocol. (Actually, the
$|\phi_{\text{b}i}\rangle$ entanglement step is often mistakenly
described as an honest one.) These possibilities have not been accounted
for. In the discussions of a proper framework for QBC protocols in
Ref.~\refe{yue5}, we have codified some intuitively valid rules for protocol formation under the names {\em Intent Principle} and {\em Libertarian
Principle}. In the following, we will discuss several of the many gaps
in the impossibility proof.

\section{No impossibility theorem without QBC definition}

A plausible first reaction to the impossibility proof is: why are all
possible QBC protocols covered by its formulation? More precisely, how
may one define the necessary feature of an unconditionally secure QBC
protocol that is required for any proof of a mathematical theorem
that says such protocol is impossible? {\em No} such definition is
available. The situation is similar to the lack of a definition of an
``effectively computable'' function in the context of the
Church-Turing thesis. Nobody calls the Church-Turing thesis the
Church-Turing theorem. This is because there is no mathematical
definition of an effectively computable
function. The logical possibility is open that someday a procedure
may be found that is intuitively or even physically effective, but which can
compute a nonrecursive arithmetical function.
\par Thus in the absence of a precise definition of a QBC protocol,
one would have at best an ``impossibility thesis'', not an
impossibility {\em theorem}. (This view was emphasized to the author
by Masanao Ozawa.) Just as there appear to be many different forms of
effective procedures, there are many different QBC protocol
types\cite{yue5} that appear not to be captured by the impossibility
proof formulation. To uphold just an ``impossibility thesis'', one
would need to prove that unconditionally secure QBC is impossible in
each of these types.

\section{Unknown versus random parameter}

The impossibility proof regards any unknown number to one party as a
random variable with a known probability distribution, from which the
purification (\ref{Psi}) may be formed. However, as it is
well-known in classical statistics, not every unknown parameter is a
random variable. In the present situation, there is an infinite number
of open possibilities, such as the number of
states and operations available, that admits no uniform probability distribution or
actual entanglement for the purpose of EPR cheats. Furthermore,
there is simply no ensemble here for the unknown parameter to be
averaged over. In an analogous situation in the quantum information
literature, this error has been recently called the ``Partition Ensemble
Fallacy Fallacy''\cite{nb}. More significantly, there is no need for
Adam to know the probability $\{\lambda_k\}$ under concealing for
every $\{\lambda_k\}$. The proper approach is to regard the state
$|\Psi\rangle$ of (\ref{Psi}) as an unknown ``parameter'' in  an
infinite space. The other
party does not need to know it, or to know its probability
distribution {\it even} if it has one, because of the following {\em Secrecy
Principle} which is a corollary of the Intent Principle and
Libertarian Principle.
\begin{itemize}
\item[] {\em Secrecy Principle}: A party does not need to reveal a secret parameter chosen by her
in whatever manner if it does not affect the security of the other
party, who cannot reject the protocol on such a basis.
\end{itemize}
Thus, generation of the secret parameter can be automatized by one
party, and it can be kept secret just as Adam can keep his bit $\sb$
secret or a secret key can be kept secret in standard cryptography.

\par Indeed, with the use of (\ref{Psi}) by Babe, it is not sufficient
for concealing to assume that one fixed $\ket{\Psi}$ is used by
her as done in the impossibility proof. Two examples are given in
Ref.~\refe{yue4}, which show that Babe can cheat by using another $\{
\lambda_k\}$ or $|\Psi\rangle$ than the one prescribed, and nothing in the
impossibility proof formulation prevents her from doing that.
If one imposes the condition that the protocol is
$\epsilon$-concealing for every possible choice of $|\Psi\rangle$,
 then there is {\em no} impossibility proof until one shows that there
is a cheating transformation for Adam which will work for every
possible $|\Psi\rangle$. In the case of perfect concealing, this has
been proved\cite{yue4} for a single use of (\ref{Psi}) by Babe. The
corresponding $\epsilon$-concealing case is yet to be resolved. See the article by
G. M. D'Ariano in this volume for a quantitative discussion.

\par Note that the Secrecy Principle directly contradicts the claim
that a pure $|\Phi_\sb\rangle$ is openly known at the end of
commitment. One consequence is that because Babe does not know
$\{p_{\sb i}\}$, the usual specification of the concealing condition is a
sufficient but not necessary one needed for a general impossibility
proof.  Furthermore, one has to show that whatever information  Adam lacks on
$|\Phi_\sb\rangle$, such as the $\ket{f_k}$ of (\ref{Psi}), is {\it not} needed for his
cheating. Observe also that (\ref{Psi}) is not equivalent to the mere
generation of $|\psi_k\rangle$ with probability $\lambda_k$, due to
the presence of off-diagonal terms $|f_k\rangle\langle f_{k'}|$.
Such purification has to be considered because of possible
entanglement cheating, not because of the Church of the Larger Hilbert
Space. Indeed, entanglement may help determine the bit through such
terms, as the example in the next section shows.   Even with the
Church, the two cases are not equivalent.

\section{Shifting of the evidence state space}  

Even when a pure $|\Phi_\sb\rangle$ is openly known, the impossibility
proof does not cover the situations in which opening and verification
are more elaborate, involving component parts of ${\cal H}^A$
and  ${\cal H}^B$. In particular, consider a protocol in which Babe
forms (\ref{Psi}) and sends Adam $\cH^{B_1}$, with $\ket{\psi_k} =
\ket{\psi_{k1}}\ket{\psi_{k2}}$ in $\cH^{B_1} =
\cH^{B_{11}}\otimes \cH^{B_{12}}$.  Adam randomly
switches the state in $\cH^{B_{11}}$ to be that of $\ket{\psi_{k1}}$ or
$\ket{\psi_{k2}}$  by the unitary perumation $P_m$,
$m \in \{1,2\}$, modulates the resulting state in $\cH^{B_{11}}$ by
a single $U_\sb$ for each $\sb$, and sends it to Babe.  He opens by
revealing $\sb$, his random permuation $P_m$, and {\it returning}
$\cH^{B_{12}}$.  Babe verifies by testing the
apropriate states in $\cH^{B_{11}}$ for checking $\sb$, and
$\cH^{B_{12}}$ for checking that there is no change.  It is possible that the
protocol is both concealing and binding for the following reason. For the final committed state $\ket{\Phi_\sb}$ with Adam
entangling the $P_m$ with $\ket{e_i} \in \cH^{A_1}$, we have $\cH^A =
\cH^{A_1} \otimes \cH^{B_{12}}$ and $\cH^B =
\cH^{B_{11}} \otimes \cH^{B_2}$.  Thus, $\rho^B_0$ can be close to
$\rho^B_1$ because $\cH^{B_{12}}$ is not
available to Babe for her cheating.  However, only $\cH^{A_1}$, and
not $\cH^A$, is avaiable to Adam's cheating, so he cannot apply the
required cheating $U^A$ without being found cheating with a
nonvanishing probability.  Using the upper bound in 
(\ref{eq:fidbound}) the security condition can be expressed as
$\rho_0^B({\cal H}^{B_{12}}\otimes{\cal H}^{B_2})\sim
\rho_1^B({\cal H}^{B_{12}}\otimes{\cal H}^{B_2})$ and 
$\rho_0^B({\cal H}^{B_1}\otimes{\cal H}^{B_2})\not\sim
\rho_1^B({\cal H}^{B_1}\otimes{\cal H}^{B_2})$. To preserve the impossibility
proof one would need to show that, in addition to (\ref{eq:ip}), $\lim_n \bar{P}_c^B({\cal H}^{B_{12}}\otimes{\cal
H}^{B_2})=\frac{1}{2}\Rightarrow \lim_n\bar{P}_c^B
(\cH^{B_1} \otimes \cH^{B_2})=\frac{1}{2}.$ Clearly, this has {\it not} been proved.

As an example, consider
the case $\cH^{B_1} = \cH^{B_{11}}
\otimes \cH^{B_{12}} \otimes \cH^{B_{13}} \otimes \cH^{B_{14}}$ of
four qubits, with $\{ \ket{\psi_k} \} =
\{\ket{1}\ket{2}\ket{3}\ket{4}, \ket{4}\ket{1}\ket{2}\ket{3},
\ket{3}\ket{4}\ket{1}\ket{2}, \ket{2}\ket{3}\ket{4}\ket{1}\}$, where
$\{ \ket{1},\ket{2},\ket{3},\ket{4}\}$ are, e.g., a fixed set $S_0$ of four
possible BB84 states on a given great circle of a qubit.  Adam
permutes each $\ket{\psi_k}$ by one of four possible $P_m$, and returns
the first qubit to Babe unchanged for $\sb=0$, while shifted by $\pi$
in the great circie for $\sb=1$.  Assume first that Babe either did
not entangle, or cannot use
her entanglement in $\cH^{B_2}$. Then $\rho^{B_{11}}_0(\psi_k) = \rho^{B_{11}}_1(\psi_k)$ for all $k$,
and no entanglement of permutations would
produce a rotation on the first qubit while not disturbing the
others.  Thus, Adam cannot cheat perfectly and has a fixed
$\bar{P}^A_c$ for this protocol which is not arbitrarily close to one,
even though it is perfectly concealing. If one can find a case in
which the protocol remains perfectly concealing with entanglement by
Babe, which is not the case in this example, (IP) of (\ref{eq:ip}) would be contradicted, and the case can be
extended to become an unconditionally secure protocol by repeating it
in a sequence. Such a case can indeed be found in this kind of
protocols which we call Type 2.

\section{Protocol QBC1}

If carried out honestly, this protocol is conceptually simple and works as follows.\cite{yue5} Adam
sends Babe $n$ qubits with states selected randomly and independently
from $S_0$.  Babe then picks randomly one of these qubits and sends it
back to Adam, who would leave it unchanged or shift it by $\pi$,
depending on whether $\sb = 0$ or 1, and commit it as evidence.  He
opens by revealing $\sb$ and all the qubit states, and Babe verifies
by corresponding measurements.

We assume that no cheating by either party, other than entanglement,
occurs during commitment as in the
impossibility proof formulation, say, under heavy penalty in a
game-theoretic formulation where state checking is done by both parties. Thus the protocol is perfectly
concealing.  There are many ways for Babe to randomly pick one of the
$n$ qubits, say by permutation into a fixed qubit among the $n$ ones,
or into a separate fixed qubit, each with its own purification.  If
Adam knows which particular way Babe chooses, it can be shown that he
can cheat successfully.  However, his success depends crucially on this
knowledge, and no further entanglement purification by Babe is possible
over these different ways that would allow her to send back a single
qubit to Adam for bit modulation.  While the situation here has some
similarity to our Type 3 protocols,\cite{yue5}, it is one that cannot
be completely purified even with a known probability distribution, and the impossibility proof does
not apply. Thus, the protocol becomes $\epsilon$-binding for large
$n$. A full security proof of this protocol and detailed treatment of
Type 2 protocols will be presented elsewhere.

\section*{Acknowledgements}

I would like to thank G.M. D'Ariano, M. Ozawa, and I.L. Chuang for useful
discussions. This work was supported by the Defense Advanced Research
Projects Agency and the Army Research Office.

\end{document}